\documentclass[proceedings, preprint]{rmaa}

\usepackage{paralist}

\usepackage{psfrag,color}

\SetYear{2023}
\SetConfTitle{III Workshop Astronomy for Inclusion}

\title{Sound training platform applied to astronomy} 

\author{
  N. M. M. Bertaina Lucero,\altaffilmark{1,2} 
  J. Casado,\altaffilmark{1}
  B. García,\altaffilmark{2,3}
  and G. Cayo\altaffilmark{3}}

\altaffiltext{1}{Instituto de Bioingenier\'\i{}a,
  Universidad de Mendoza, Ciudad de Mendoza, Argentina.}

\altaffiltext{2}{Consejo Nacional de Investigaciones Científicas y Técnicas - CONICET.}

\altaffiltext{3}{Universidad Tecnológica Nacional, Facultad Regional Mendoza, Ciudad de Mendoza, Argentina.}

\shortauthor{Bertaina Lucero, Casado, García \& Cayo}
\shorttitle{RevMexAA(SC) Sound training platform}

\listofauthors{N. M. M. Bertaina Lucero, J. Casado, B. García, \& G. Cayo}
\indexauthor{Bertaina Lucero, N. M. M.}
\indexauthor{Casado, J.}
\indexauthor{García, B.}
\indexauthor{Cayo, G.}

\abstract{The convergence between astronomy and data sonification represents a significant advancement in the approach and analysis of cosmic information. By surpassing the visual exclusivity in data analysis in astronomy, innovative projects have developed software that goes beyond visual representation, transforming data into auditory and tactile displays. However, it has been evidenced that this novel technique requires specialized training, particularly for audio format data. This work describes the initial development of a platform aimed at providing training for data analysis in astronomy through sonification. The integration of these tools in astronomical education and research opens new horizons, facilitating a more inclusive and multisensory participation in the exploration of space science.}

\resumen{La convergencia entre la astronomía y la sonorización de datos marca un avance significativo en la forma en que se aborda y analiza la información del cosmos. Superando la exclusividad visual en el análisis de datos en astronomía, proyectos innovadores han desarrollado software que va más allá de la representación visual, transformando datos en despliegues sonoros y táctiles. Sin embargo, se ha evidenciado que esta novedosa técnica requiere un entrenamiento especializado, particularmente para datos en formato de audio. En este trabajo se describe el desarrollo inicial de una plataforma que tiene como objetivo principal contener entrenamientos para el análisis de datos en astronomía a través de la sonorización. Lograr la integración de estas herramientas en la educación y la investigación astronómica abre nuevos horizontes, facilitando una participación más inclusiva y multisensorial en la exploración de las ciencias del espacio.}

\addkeyword{Sonorización de datos}
\addkeyword{Entrenamientos}
\addkeyword{Accesibilidad}
\addkeyword{Multisensorialidad}

\begin{document}
\maketitle

\section{Introduction}
\label{sec:intro}
When discussing astronomy, the mind often associates it with numbers, images of celestial bodies, and data in various formats, but it is always presented in tables for visual inspection. This predominance of visual expressions in astronomy can be challenging for people with visual impairments. However, there are projects aimed at promoting inclusion in this field. These projects have led to the development of software that goes beyond visual representation of data, transforming it into sound and even tactile information. Some of the most relevant projects to date with active support include:  HighCharts Sonification Studio \citep{2017Proceedings}, sonoUno \citep{2019SonoUno} and StarSound \citep{2022Star}. These projects offer a user interface that makes them more accessible to people without programming knowledge.

On the other hand, some packages allow for the sonification of data without a user interface, such as Strauss \citep{2021Harrison}, Astronify\footnote{https://astronify.readthedocs.io/en/latest/} and Soni-py \citep{2021Cambridge}. In addition to being used for the analysis of astronomical data, all of these tools can be applied to sonify data of various types, allowing for scientific research and development in a wide range of areas.

With the use of this innovative approach to data interpretation in recent years, the need for specialized training has become evident so that users can approach, understand, and analyze data through sonification \citep{2024Proceedings}. Although the sense of hearing is used from the moment a person is born, the study and subsequent use of a new technique always requires training that indicates exactly what is being represented and how it should be interpreted.

Most of the programs available for auditory training with visual support have focused on areas related to language and its comprehension. Among the diversity of these, could be found some such as that of \citet{2023American_journal}, which summarizes the training with visual-auditory biofeedback that is used to improve speech in individuals with difficulties detecting these sounds. However, in the search for more innovative applications, some proposals sonify numerical data to explore the relationship between sound information and the understanding of visual representations, determining whether the sonification of graphics can improve understanding, especially in the educational field, and the identification of features in signals, over noise, in the case of scientific research \citet{2001GeorgiaInstitute}.

Following this line of research, the early development of a web platform for training in sonification of different types of astronomical data will be described. These programs can not only help understand sound perception but also offer the opportunity to integrate this novel tool into data analysis that would ultimately allow more people to engage in science, harnessing multiple senses simultaneously to explore nature.

\section{Preevious works}
\label{sec:estado}

\begin{figure}[!t]
  \includegraphics[width=\columnwidth]{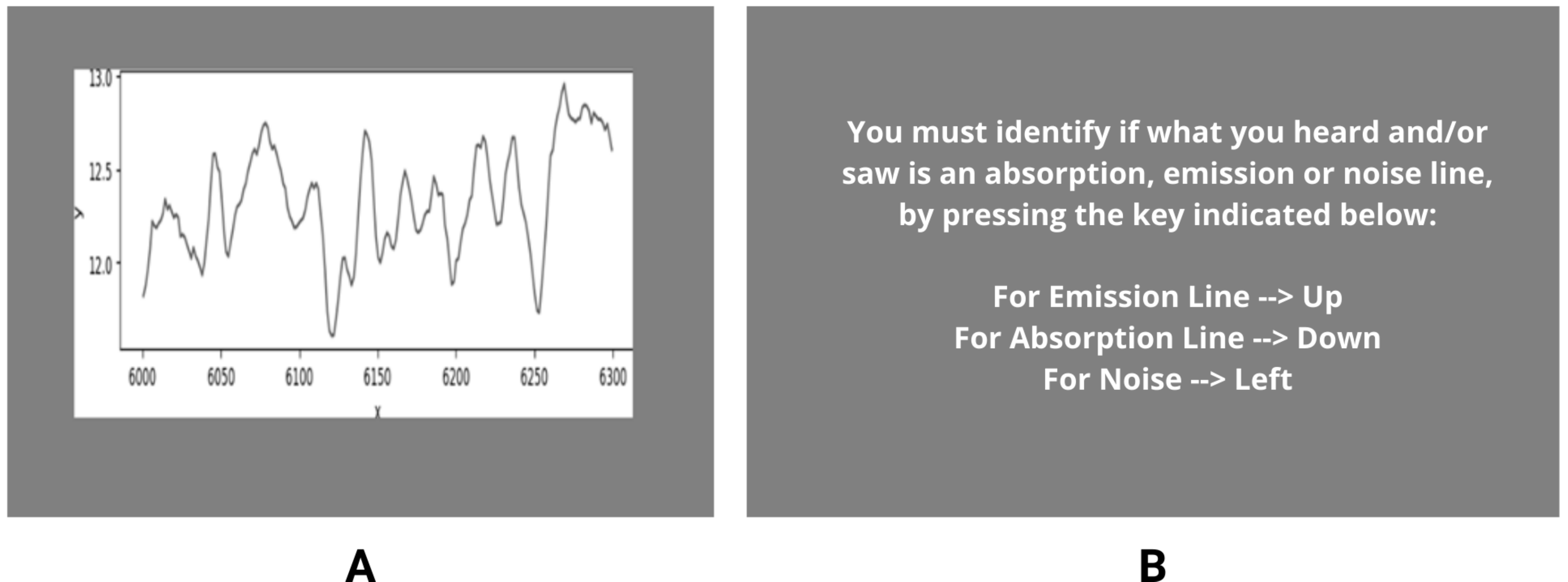}
  \caption{A. Image and sound display in the training designed with Psychopy. B. Question and options are displayed for the user to answer using the Arrow Key block of the keyboard.}
  \label{fig:1}
\end{figure}

In 2022, \citet{2023Arvix} conducted a series of trainings using the Psychopy tool \citep{2022Book}, which is primarily useful for conducting experiments in behavioral sciences such as neuroscience, linguistics, and psychology, among others. In the first proposed training, specific spectral data such as emission and absorption lines were used, which were displayed both graphically and sonorously so that the participant could classify whether what they heard and saw was one of the two mentioned signals (Fig.\ref{fig:1}). In addition to a block of visual and auditory display, another block with auditory display only was included to test whether visual support should be included at the beginning of the training \citep{2024Proceedings}.

In the second stage of development, a training workshop was conducted over two sessions of 2 hours each. Different blocks or work units were programmed for this workshop, with the sonification of various types of data, such as noise produced in the detector of the Gravitational Wave Observatory (EGO) (Glitches), particles from the Large Hadron Collider (LHC), and cosmic muons used for muongraphy. Those data were extracted from Casado's doctoral thesis\citep{2023Casado}. These tools and techniques can be applied to various fields, including astronomy, and can help improve data interpretation and analysis skills.

The mentioned workshop can be considered a focus group that allowed for a first approach to the use of the tool and to collect user feedback. This is how the need for training on a dedicated website arose, as Psychopy is a desktop program that offers a website for executing the experiments but has the disadvantage of being paid. In addition, during these meetings, suggestions for improvement were obtained from users, which were considered for the original, open, and free web deployment.

\section{Methods and Tools}
\label{sec:web}

Taking into account the existing tools for developing a website, ranging from designing the visible part for the user (front-end) to the part that works with the server (back-end) and programming the database for collecting user responses, it is proposed to use the Django web development framewor\footnote{2005-2023 Django Software Foundation and individual contributors. Available at: https://www.djangoproject.com/}. It is a high-level open-source web framework with active support to date. It offers extensive documentation, which is up-to-date with the most current version to date (v5.0). It allows the creation of complex websites, programming most of its parts in Python, standing out for its effective Model-View-Tempered architecture. Each of these components is used to handle different parts of a website. On the one hand, the models generate structures that represent different tables in the database used and help in its management and manipulation. The views will be in charge of handling the interactions of the models and templates that the user sees. Finally, templates are HTML-type files that represent what will be shown to the user. As for the database, Postgresql was used \footnote{1996-2023 The PostgreSQL Global Development Group. Available at: https://www.postgresql.org/}. Finally, some functionalities of Bootstrap 5 framework were integrated \footnote{Bootstrap team. Available at: https://getbootstrap.com/}, which includes design and utilities in a few lines of code that make the development of a clean and organized front-end more fluid, both at the code and user interface level. Additionally, it integrates classes in the tags that allow for responsive design (adapts to different screen sizes).

\section{Development}
\label{sec:desarrollo}

\begin{figure}[!t]
  \includegraphics[width=\columnwidth]{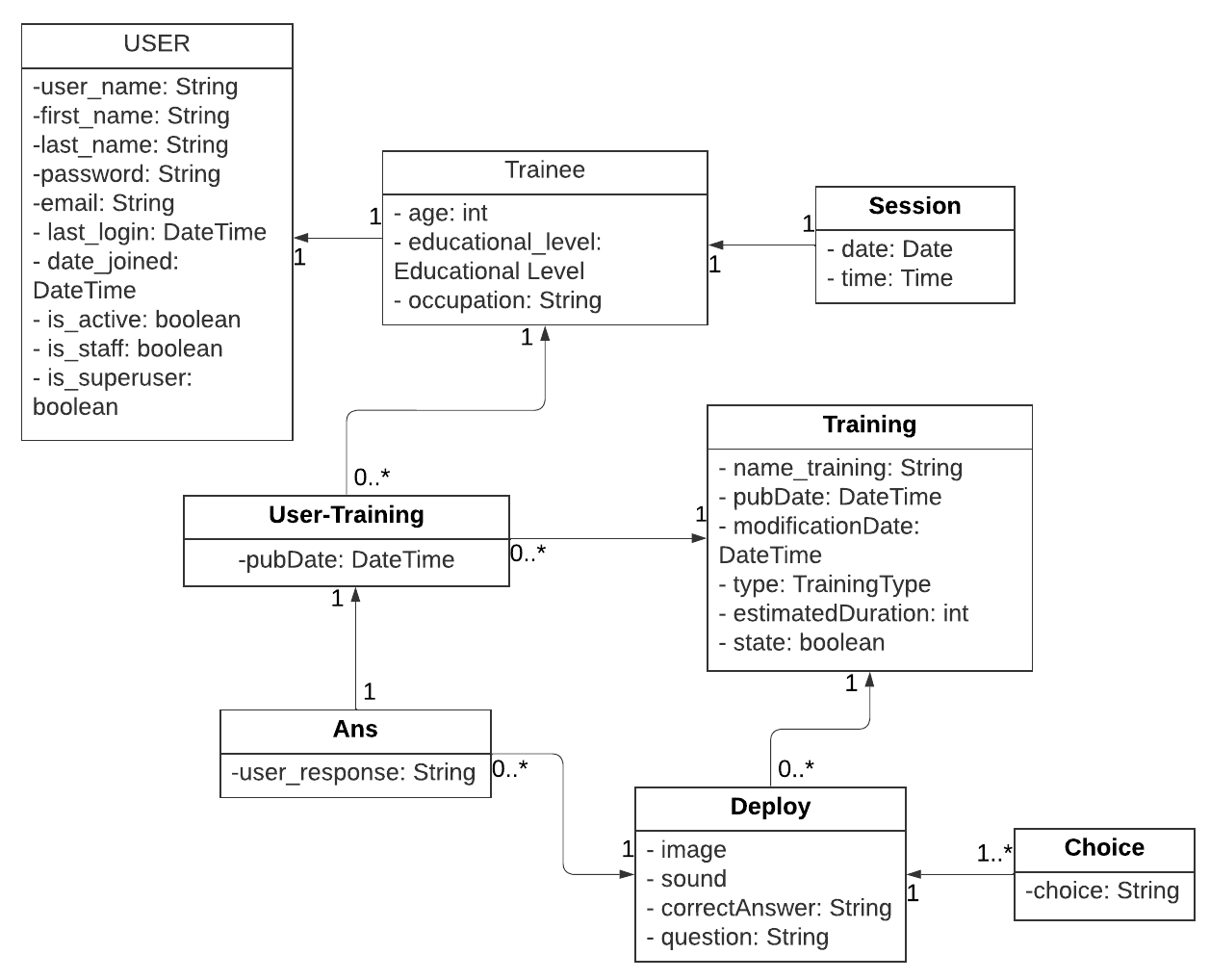}
  \caption{Basic diagram of the design of the database, represented and used in Django models.}
  \label{fig:2}
\end{figure}

The website was completely implemented using Django, which allowed the programming of user views using HTML and CSS files, as well as the management of the database exclusively with the Python language. Concerning the database, among the various management systems available, PostgreSQL was chosen due to its easy integration with the Django project, providing an effective connection between the developed application and the programmed database. In addition to this, a database made with PostgreSQL stands out for its ability to manage large volumes of data, which is crucial for the scalability of the site in the future.
The database not only provides support for the management of the data received when the user finishes the different training sessions but also helps to manage all the files that will be displayed, both visually and by sonification, as well as the user information for the start session (Fig. \ref{fig:2}).

The entire database is controlled through an administration view that helps obtain the necessary statistics to adjust the different training programs that are developed. From these same statistics, it is possible to have a report on the general progress of the users, asking them to adjust the programs as required. In addition, with the data obtained from the number of correct answers given, the user can be kept informed of their progress. In this same database, responses to small surveys conducted with users will be stored, which will contribute to the continuous improvement of training programs.

Regarding the front-end construction of the site, attention was paid to the appropriate use of tags and an effective distribution of all its sections. This was mainly guided by the goal of making the site accessible to as many users as possible. Although the process is subject to changes with the changing versions of the languages used in its construction, it is aimed to achieve the highest accessibility percentage possible, although it cannot be guaranteed at a hundred percent.

\begin{figure}[!t]
  \includegraphics[width=\columnwidth]{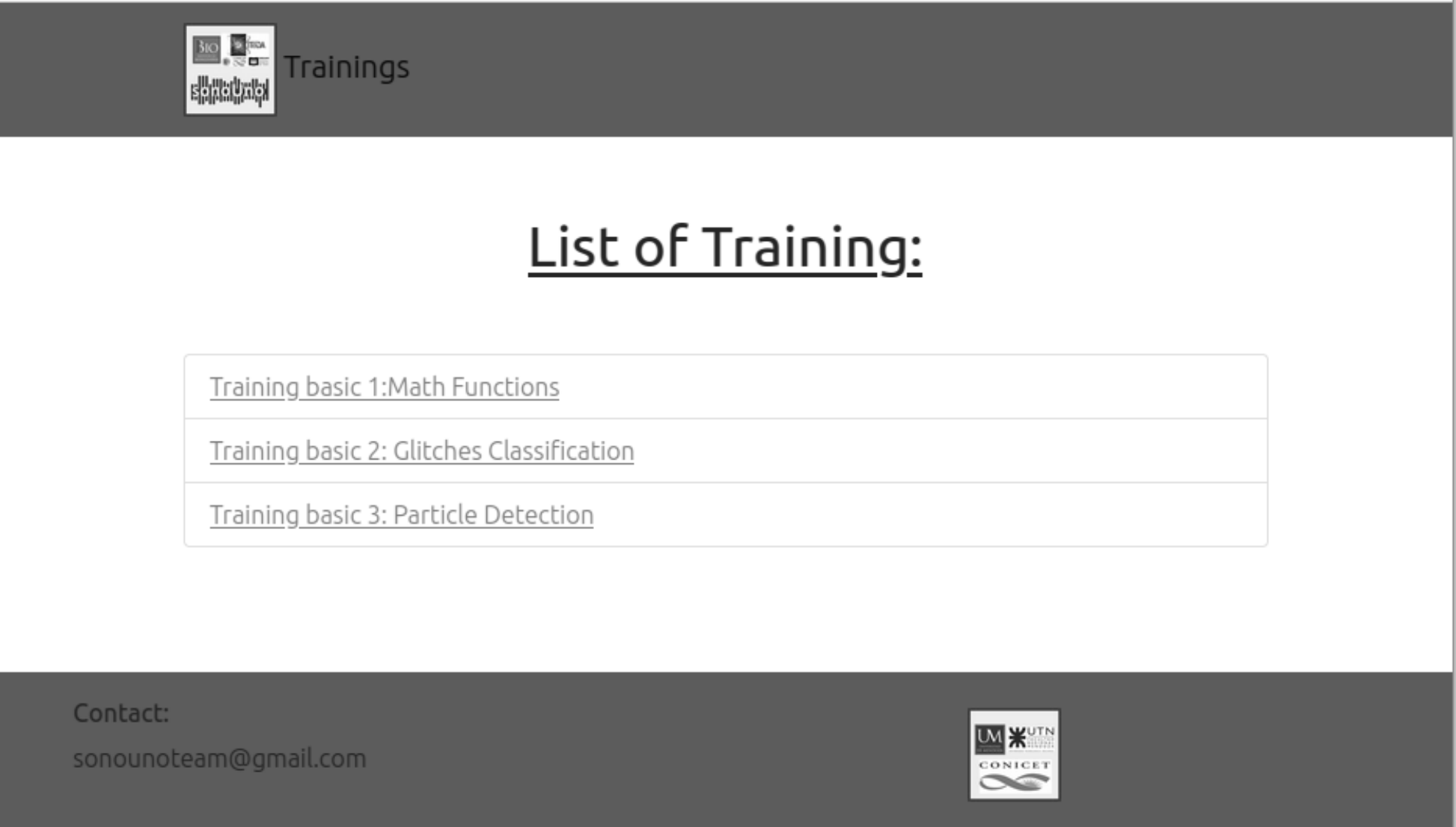}
  \caption{Training page, with the list corresponding to the beginner level.}
  \label{fig:3}
\end{figure}

The platform designed for training is intended to be connected to the sonoUno web, providing a better understanding of sonified data. It is planned to track each user's progress, classifying them into three categories: beginner, intermediate, and advanced (Fig.\ref{fig:3}). Depending on the category in which a user is, they will have access to a series of training adapted to their level. As users correctly complete the training, they will be promoted to a higher category, unlocking new training with increasing levels of difficulty.

\begin{figure}[!t]
  \includegraphics[width=\columnwidth]{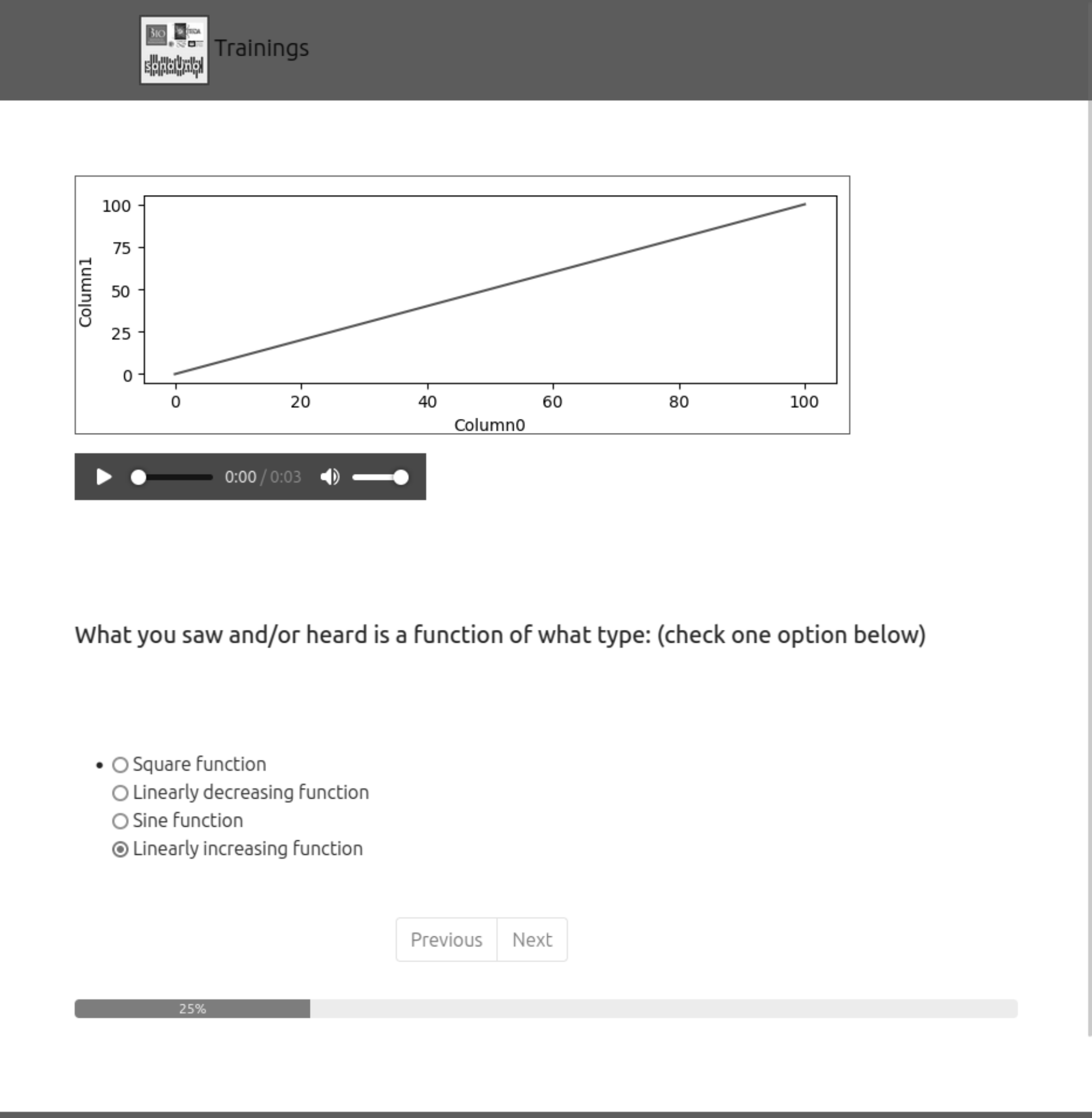}
  \caption{Example of deployment of mathematical functions training: beginner level.}
  \label{fig:4}
\end{figure}

Each training is organized into blocks, which are further divided into four displays arranged in a form. Each of these latter consists of presenting an image, audio, and a question with a series of options for the user to answer, maintaining a paginated overall view (see Fig.\ref{fig:4}). This structure helps to prevent the user from having to scroll down to interact with the next display, making the experience more comfortable and avoiding adding additional cognitive load.

Regarding the images and sounds used in the development of the training, these were obtained from previously selected real astronomical data using sonoUno. The only exception to this characteristic is found in the initial training intended for all users, which consists of presenting simple mathematical functions (Fig.\ref{fig:4}), serving as an introduction to the training methodology.

It is important to clarify about sonified data, that it is not mandatory to exclusively use sonoUno. Any of the sonification programs mentioned earlier can be used. This makes the platform versatile for training users in data sonification in general, regardless of the software used for this purpose.

The final website is in the development stage. For it to move to the production stage, some accessibility tests must be completed, such as testing the deployment with screen readers. Ensuring accessibility and inclusion is considered a fundamental pillar of this development.
 
\section{Conclusion}
\label{sec:conclusion}

Sonification of different types of data represents an advance in the way of approaching research, as well as can create the path towards greater inclusion in science. This is why the introduction of different software that translates data into sound, visual, and tactile displays not only challenges accessibility barriers but also offers a new perspective for exploring and analyzing information.

The development of specific training programs is essential to unlocking the full potential of this innovative approach. In addition to this, a specific educational approach is required to be able to interpret the subtleties found in this new display of information, which mainly involves discriminating a signal from noise that may be natural or linked to instrumental problems. This task is not trivial; it requires an exhaustive study, which allows topics such as those linked to perception to be included in the results.

On this path to providing an optimal training platform, it is important to choose a framework that allows a comprehensive approach, from the user interface to database management, as allowed by Django. Although the website is still in development, its main goal is to provide adaptive training that allows an innovative educational experience for the interpretation of sonified data.


\begin{thebibliography}

\bibitem[Bertaina Lucero(2023)]{2023Arvix} Bertaina Lucero, N. M. M. \ 2023, in: ArXiv, abs/2305.06943.
\bibitem[Bonebright et al.(2001)]{2001GeorgiaInstitute} Bonebright, T. L., Nees, M. A., Connerley, T. T., \& McCain, G. R. \ 2001, in:
    \textit{Testing the effectiveness of sonified graphs for education: A programmatic research project.} (Proceedings of the 2001 International Conference on Auditory Display)
\bibitem[Casado(2023)]{2023Casado} Casado, J. \ 2023, in: https://arxiv.org/pdf/2305.05635.pdf
\bibitem[Casado et al.(2019)]{2019SonoUno} Casado, J., De La Vega, G., Díaz-Merced, W., Gandhi, P., \& García, B. \ 2019 in:
    \textit{SonoUno: a user-centred approach to sonification}. Proceedings of the International Astronomical Union. 2019;15(S367):120-123. doi:10.1017/S174392132100079X
\bibitem[De la Vega et al.(2023)]{2023Web} De la Vega, G., Dominguez, L., Casado, J., \& García, B. \ 2023, in: SonoUno web: an innovative user centred web interface. (10.48550/arXiv.2302.00081)
\bibitem[Foran et al.(2022)]{2022Star} Foran, G., Cooke, J., \& Hannam, J. \ 2022, Revista Mexicana de Astronom\'\i{}a y Astrof\'\i{}sica Serie de Conferencias, 54, 1-8.
\bibitem[Harrison et al.(2021)]{2021Harrison} Harrison, C., Trayford, J., Harrison, L., \& Bonne, N. \ 2021, in:
    \textit{Audio universe tour of the solar system: using sound to make the universe more accessible.} Astronomy and Geophysics, 63(2):2.38–2.40.
\bibitem[Hitchcock et al.(2023)]{2023American_journal} Hitchcock, E. R., Ochs, L. C., Swartz, M. T., Leece, M. C., Preston, J. L., \& McAllister, T. \ 2023, in:
    \textit{Tutorial: Using Visual–Acoustic Biofeedback for Speech Sound Training.} American journal of speech-language pathology, 32(1), 18-36.
\bibitem[Kondak et al.(2017)]{2017Proceedings} Kondak, Z., Liang, T. A., and Tomlinson, B., \& Walker, B. N. \ 2017, in: 
    \textit{Web sonification sandbox-an easy-to-use web application for sonifying data and equations.} In Proceedings of 3rd Web Audio Conference, London.
\bibitem[Patton and Levesque(2021)]{2021Cambridge} Patton, L., \& Levesque, E. \ 2021, in GitHub (soni-py: A pitch-based data sonification package.) Department of Astronomy, University of Washington, Seattle; Harvard-Smithsonian Center for Astrophysics, Cambridge.
\bibitem[Lopez and Tello(2024)]{2024Proceedings} Lopez, N. M., \& Tello, E. \ 2024 in: 
    \textit{Advances in Bioengineering and Clinical Engineering.} Proceedings of the XXIII Argentinian Congress of Bioengineering (SABI 2022) and the XII Clinical Engineering Conference. 
\bibitem[Peirce et al.(2022)]{2022Book} Peirce, J. W., Hirst, R. J., \& MacAskill, M. R. \ 2022. Building Experiments in PsychoPy. 2nd Edn London: Sage.

\end{thebibliography}
\end{document}